\documentclass[aps,prd,preprint,superscriptaddress,showpacs]{revtex4-1}
\usepackage{amssymb,amsmath}
\usepackage{graphicx} % Include figure files
\usepackage{dcolumn}  % Align table columns on decimal point
\usepackage{feynmf}   % Drawing Feynman Diagrams
\usepackage{epsfig}   % For PostScript figures
\newcommand{\tr}{ {\mathbf{tr}\, }}
\newcommand{\Tr}{ {\mathbf{Tr}\, }}

\begin{document}

% Use the \preprint command to place your local institutional report
% number in the upper righthand corner of the title page in preprint mode.
% Multiple \preprint commands are allowed.
% Use the 'preprintnumbers' class option to override journal defaults
% to display numbers if necessary
\preprint{}

%Title of paper
\title{On QCD and Effective Locality}

% repeat the \author .. \affiliation  etc. as needed
% \email, \thanks, \homepage, \altaffiliation all apply to the current
% author. Explanatory text should go in the []'s, actual e-mail
% address or url should go in the {}'s for \email and \homepage.
% Please use the appropriate macro foreach each type of information

% \affiliation command applies to all authors since the last
% \affiliation command. The \affiliation command should follow the
% other information
% \affiliation can be followed by \email, \homepage, \thanks as well.
\author{H. M. Fried}
\thanks{Supported in part by a Julian Schwinger Foundation Travel Grant.}
\affiliation{Physics Department, Brown University, Providence, RI 02912, USA}
\author{M. Gattobigio}
\affiliation{Universit\'{e} de Nice Sophia-Antipolis,\\ Institut Non Lin$\acute{e}$aire de Nice, UMR 6618 CNRS, 06560 Valbonne, France}
\author{T. Grandou}
\affiliation{Universit\'{e} de Nice Sophia-Antipolis,\\ Institut Non Lin$\acute{e}$aire de Nice, UMR 6618 CNRS, 06560 Valbonne, France}
\author{Y.-M. Sheu}
\email{ymsheu@mailaps.org}
\affiliation{Universit\'{e} de Nice Sophia-Antipolis,\\ Institut Non Lin$\acute{e}$aire de Nice, UMR 6618 CNRS, 06560 Valbonne, France}

%\homepage[]{Your web page}
%\thanks{}
%\altaffiliation{}
%\affiliation{Institut Non Lin\'{e}aire de Nice Sophia-Antipolis, UMR-CNRS 6618}
%Collaboration name if desired (requires use of superscriptaddress
%option in \documentclass). \noaffiliation is required (may also be
%used with the \author command).
%\collaboration can be followed by \email, \homepage, \thanks as well.
%\collaboration{}
%\noaffiliation

\date{\today}

\begin{abstract}
In a recent paper it was shown how quark scattering in a quenched, eikonal model led to a momentum-transfer dependent amplitude expressed in terms of Halpern's functional integral; and how the requirement of manifest gauge invariance converted that functional integral into a local integral, capable of being evaluated with precision by a finite set of numerical integrations.  We here prove that this property of "effective locality" holds true for all quark processes, without approximation and without exception.
\end{abstract}

% insert suggested PACS numbers in braces on next line
%\pacs{12.38.-t}{Quantum chromodynamics}
%\pacs{12.38.Lg}{Other nonperturbative calculations }
%\pacs{11.15.-q}{Gauge field theories}
% insert suggested keywords - APS authors don't need to do this
%\keywords{}

%\maketitle must follow title, authors, abstract, \pacs, and \keywords
\maketitle

% body of paper here - Use proper section commands
% References should be done using the \cite, \ref, and \label commands

%
%
%
\section{Introduction}

In a recent paper~\cite{Fried2009} it was shown how the requirement of manifest gauge invariance of the Schwinger functional solution of QCD leads to a description of all possible virtual gluons exchanged between quarks as "ghosts", wherein the form of the final, Halpern~\cite{Halpern1977a,Halpern1977b,Halpern1979a} functional integral becomes "effectively local".  The resulting amplitudes contain all possible gluon exchanges, including cubic and quartic gluon interactions, and are rendered gauge invariant by becoming independent of exchanged gluon propagators.

What takes the place of virtual boson propagator-exchange in an Abelian theory, is, in QCD, an effectively-local version of the Halpern variable, $\chi^{a}_{\mu \nu}$, with the huge simplifications resulting from the replacement of the Halpern functional integral by a small set of ordinary integrals. Even before such integrals were evaluated, one could see in Ref.~\cite{Fried2009} the possible underlying structure of asymptotic freedom and the  MIT Bag Model.  That discussion was carried out in the context of a high-energy eikonal model, in quenched approximation.  We prove below that similar, "effectively local" structures will be obtained in the evaluation of the exact theory; the final integrands will surely be more complicated than those of Ref.~\cite{Fried2009}, but they will always appear as a finite set of integrals which can be evaluated numerically or approximately.

The article is organized as follows. Section~\ref{Sec:2} summarizes the main steps of the QCD generating functional's derivation using functional methods, and, in particular, the so-called `linkage operator'. Halpern's trick, to re-formulate the gluonic action, is used here as an important step of our derivations. Section~\ref{Sec:3} gives the explicit Fradkin representations for the Green's function of a quark propagating in an external gauge field, and for the quark closed-loop functional $\mathbf{L}[A]$. For additional clarity, Section~\ref{Sec:3} is supported by an Appendix.  The Effective Locality property is presented in Section~\ref{Sec:4}, as well as some comments concerning its meaning as well as its power in the course of practical calculations.  A brief discussion and summary are presented in Section~\ref{Sec:5}.

\section{\label{Sec:2}QCD Generating Functional}

We begin with the QCD Lagrangian
\begin{equation}\label{Eq:QCDLagrangian01}
\mathcal{L}_{\mathrm{QCD}} = - \frac{1}{4} \mathbf{F}_{\mu \nu}^{a} \mathbf{F}_{\mu \nu}^{a} - \bar{\psi} \cdot [m + \gamma_{\mu} \, (\partial_{\mu} - i g A_{\mu}^{a} \, \tau^{a})] \cdot \psi,
\end{equation}

\noindent where $\mathbf{F}_{\mu \nu}^{a} = \partial_{\mu} A_{\nu}^{a} -  \partial_{\nu} A_{\mu}^{a} + g f^{abc} A_{\mu}^{b} A_{\nu}^{c}$ is the field strength and $\tau^{a}$'s are the color matrices of SU(N) with $[\tau^{a}, \tau^{b}] = i f^{abc} \tau^{c}$.  An `economic way' to proceed consists of adding and subtracting a gauged free gluon term $\mathcal{L}^{(0)}_{\mathrm{gluon}}$ into the Lagrangian such that
\begin{subequations}\label{Eq:GluonLagrangian03}
\begin{eqnarray}
\mathcal{L}^{(0)}_{\mathrm{gluon}} &=& - \frac{1}{4} \mathbf{f}_{\mu \nu}^{a} \mathbf{f}_{\mu \nu}^{a} -  \frac{1}{2\zeta} (\partial_{\mu} A_{\mu}^{a})^{2},\\
\mathcal{L}'_{\mathrm{gluon}} &=& \mathcal{L}_{\mathrm{gluon}} - \mathcal{L}^{(0)}_{\mathrm{gluon}} = - \frac{1}{4} \mathbf{F}_{\mu \nu}^{a} \mathbf{F}_{\mu \nu}^{a} - \left(- \frac{1}{4} \mathbf{f}_{\mu \nu}^{a} \mathbf{f}_{\mu \nu}^{a} - \frac{1}{2\zeta} (\partial_{\mu} A_{\mu}^{a})^{2} \right),
\end{eqnarray}
\end{subequations}

\noindent where $\mathbf{f}_{\mu \nu}^{a} = \partial_{\mu} A_{\nu}^{a} -  \partial_{\nu} A_{\mu}^{a}$ is the field strength at zero coupling, $\mathbf{f}_{\mu \nu}^{a}=\mathbf{F}_{\mu \nu}^{a}(g=0)$, and where $\zeta$ is the gauge parameter.  Such separation provides a simple, functional definition of free and interacting gluonic Lagrangian, with the covariant gauge dependence of the free gluon propagator given by Eq.~(4), below.  The only purpose here, is to provide the gluonic $A^a_\mu(x)$-fields with a well-defined (invertible) quadratic form, which in the present example, turns out to be given by the covariant propagator, since the gauged free gluon action can effectively be written as
\begin{equation}
\frac{i}{4} \int{\mathbf{f}_{\mu \nu}^{a} \mathbf{f}_{\mu \nu}^{a}} + \frac{i}{2\zeta} \int{(\partial_{\mu} A_{\mu}^{a})^{2}} = \frac{i}{2} \, \int{A \cdot \left( \mathbf{D}_{c}^{(\zeta)} \right)^{-1} \cdot A}
\end{equation}

\noindent where
\begin{equation}
\left. \mathbf{D}_{c}^{(\zeta)} \right|^{ab}_{\mu \nu} = \delta^{ab} (-\partial^{2})^{-1} \left[ g_{\mu \nu} - \zeta \frac{\partial_{\mu} \partial_{\nu}}{\partial^{2}} \right]
\end{equation}

\noindent and where the $g_{\mu \nu}$ are coefficients of the Minkowskian metric tensor.

\par
Then the QCD generating functional can be constructed in either the Schwinger or Symanzik manner~\cite{HMF1972},
\begin{eqnarray}\label{Eq:QCDGF01}
\mathcal{Z}[j,\bar{\eta},\eta] = {\cal{N}}\exp{\left[ i \int{\mathcal{L}'_{\mathrm{QCD}}\left[\frac{1}{i} \frac{\delta}{\delta j}, \frac{1}{i} \frac{\delta}{\delta \bar{\eta}}, \frac{-1}{i} \frac{\delta}{\delta \eta} \right]} \right] } \cdot \mathcal{Z}_{0}[j,\bar{\eta},\eta],
\end{eqnarray}

\noindent where ${\cal{N}}$ is a normalization constant, $j_{\mu}^{a}$, $\eta_{\mu}$, and $\bar{\eta}_{\mu}$ are gluon, quark and anti-quark sources, respectively, and $\mathcal{Z}_{0}$ is the generating functional of the free system, so as to obtain, after a simple rearrangement,\cite{HMF1972,HMF1990}
\begin{eqnarray}\label{Eq:QCDGF03}
\mathcal{Z}[j, \eta, \bar{\eta}] &=&{\cal{N}} \left. e^{\frac{i}{2} \int{j \cdot {\mathbf{D}_{\mathrm{c}}^{(\zeta)}} \cdot j }} \, e^{\mathfrak{D}_{A}} \, e^{ i \int{\mathcal{L}'_{\mathrm{gluon}}[A]}} \, e^{i \int{ \bar{\eta} \cdot \mathbf{G}_{\mathrm{c}}[A] \cdot \eta} } \, {e^{\mathbf{L}_{\mathrm{c}}[A]}}\right|_{A = \int{\mathbf{D}_{\mathrm{c}}^{(\zeta)} \cdot j}} ,
\end{eqnarray}

\noindent where $\exp{[\mathfrak{D}_{A}]}$ with
\begin{equation}
\mathfrak{D}_{A} = - \frac{i}{2} \int {\rm{d}}^4x\int {\rm{d}}^4y{\frac{\delta}{\delta A^{a}_{\mu}(x)} \,  \mathbf{D}_{c}^{(\zeta)}{}^{ab}_{\mu \nu}(x-y) \, \frac{\delta}{\delta A^{b}_{\nu}}(y) },
\end{equation}

\noindent  denotes the linkage operator which is to act upon all the vector field $A^{a}_{\mu}(z)$ dependence contained in all terms which follow it. Likewise, $\mathbf{G}_{c}(x,y|A)$ stands for the causal Green's function of a quark in a background gauge field $A_\mu$,
\begin{equation}\label{Eq:GreensFunction01}
\mathbf{G}_{c}[A] = \mathbf{S}_{c} \left[ 1 - i g \gamma_{\mu}  \, A^{a}_{\mu} \, \tau^{a} \, \mathbf{S}_{c}\right]^{-1} = \left[ m + \gamma \cdot (\partial - i g A \cdot \tau) \right]^{-1},
\end{equation}

\noindent and $\mathbf{L}[A]$ is the logarithm of the quark determinant,
\begin{equation}\label{Eq:ClosedFermionLoop01}
\mathbf{L}[A] = \Tr{\ln{\left[ 1 - ig \, (\gamma \cdot A \cdot \tau ) \, \mathbf{S}_{c}\right]}}.
\end{equation}

Next, an important step of our derivation makes use of Halpern's familiar functional representation of the purely gluonic contribution to the action integral~\cite{Fried2009,Reinhardt1996}
\begin{equation}\label{Eq:QCDGF08_Halpern3}
e^{- \frac{i}{4} \int{\mathbf{F}_{\mu \nu}^{a} \mathbf{F}_{\mu \nu}^{a}}}  = \tilde{\mathcal{N}}_{{\chi}} \, \int{ \mathrm{d}[{\chi}]  \, e^{\frac{i}{4} \int{\chi_{\mu \nu}^{a} \chi_{\mu \nu}^{a}} + \frac{i}{2} \int{\chi_{\mu \nu}^{a}  \mathbf{F}_{\mu \nu}^{a}  } } }
\end{equation}

\noindent with the measure $\int{\mathrm{d}[{\chi}]} = \prod_{i} \prod_{a>b} \prod_{\mu \nu} \, \int{\mathrm{d}{\chi}_{\mu \nu}^{a b}(w_{i})} $, and where $\tilde{\mathcal{N}}_{{\chi}}$ is the normalization constant of the functional integral. This allows us to re-write the gluonic interaction part of Eq.~(\ref{Eq:QCDGF03}) as
\begin{eqnarray}\label{Eq:QCDGF05_gluon_extra02}
e^{ i \int{\mathcal{L}'_{\mathrm{gluon}}[A]}} = \tilde{\mathcal{N}}_{{\chi}} \, \int{ \mathrm{d}[{\chi}] \, e^{\frac{i}{4} \int{\chi_{\mu \nu}^{a} \chi_{\mu \nu}^{a}} + \frac{i}{2} \int{\chi_{\mu \nu}^{a}  \mathbf{F}_{\mu \nu}^{a}  } } } \cdot e^{\frac{i}{2} \, \int{A \cdot \left( \mathbf{D}_{c}^{(\zeta)} \right)^{-1} \cdot A} },
\end{eqnarray}

\noindent and eventually the QCD generating functional as
\begin{eqnarray}\label{Eq:1}
\mathcal{Z}[j, \eta, \bar{\eta}] &=& \mathcal{N} \int{\mathrm{d}[{\chi}] \, e^{\frac{i}{4} \int{\chi^{2}}} \cdot e^{\frac{i}{2} \int{j \cdot \mathbf{D}_{c}^{(\zeta)} \cdot j}} } \\ \nonumber & & \quad \cdot  \left. e^{\mathfrak{D}_{A}} \cdot e^{\frac{i}{2} \int{\chi \cdot \mathbf{F}} + \frac{i}{2} \int{A \cdot \left( \mathbf{D}_{c}^{(\zeta)} \right)^{-1} \cdot A} } \cdot e^{ i \int{\bar{\eta} \cdot \mathbf{G}_{c}[A] \cdot \eta}} \cdot {e^{\mathbf{L}_{\mathrm{c}}[A]}} \right|_{A = \int{\mathbf{D}_{c}^{(\zeta)} \cdot j}}.
\end{eqnarray}

The correlation functions of QCD are obtained by appropriate functional differentiation of Eq.~(\ref{Eq:1}) with respect to gluon and quark sources; and since we are here concerned only with quark ($Q$) or anti-quark ($\bar{Q}$) interactions, in which all possible number of virtual gluons are exchanged, we immediately set the gluon sources $j^{a}_{\mu}$ equal to zero.  All $Q$/$\bar{Q}$ amplitudes are then obtained by pair-wise functional differentiation of the quark sources, $\eta^{a}_{\mu}$ and $\bar{\eta}^{b}_{\nu}$; and each such operation "brings down" one of (properly anti-symmetrized) Green's functions $\mathbf{G}_{c}[A]$.  For example, the 2-point quark propagator will involve the Functional Integral $\int{\mathrm{d}[{\chi}]}$ and the linkage operator acting upon $\mathbf{G}_{c}(x, y|A) \cdot e^{\mathbf{L}[A]}$, followed by setting $A^{a}_{\mu} \rightarrow 0$.

Similarly, the $Q$/$\bar{Q}$ scattering amplitude will be obtained from the same functional operations acting upon the (anti-symmetrized) combination $\mathbf{G}_{c}[A] \cdot  \mathbf{G}_{c}[A] \cdot \exp{(\mathbf{L}[A])}$, followed by the prescription of $A \rightarrow 0$,
\begin{eqnarray}\label{Eq:4pointfunction01}
\mathbf{M}(x_{1}, y_{1}; x_{2}, y_{2}) &=& \left. \frac{\delta}{\delta \bar{\eta}(y_{1})} \cdot \frac{\delta}{\delta \bar{\eta}(y_{2})} \cdot \frac{\delta}{\delta \eta(x_{1})} \cdot \frac{\delta}{\delta \eta(x_{2})}  \mathcal{Z}[j, \eta, \bar{\eta}]  \right|_{\eta=\bar{\eta}=0; j=0} \\ \nonumber &=& \left. i^{2} \, e^{\mathfrak{D}_{A}} \, e^{ i \int{\mathcal{L}'_{\mathrm{gluon}}[A]}} \, \mathbf{G}_{\mathrm{c}}(y_{1}, x_{1}|gA) \, \mathbf{G}_{\mathrm{c}}(y_{2}, x_{2}|gA) \, e^{\mathbf{L}[A]} \right|_{A=0},
\end{eqnarray}

\noindent  and other fermionic 2n-point functions can be derived in the same way.

\section{\label{Sec:3}Fradkin's Representation of Green's function}

The property essential for further, non-perturbative analysis of all such correlation functions is the existence of Fradkin representations~\cite{Fradkin1966a,HMF1990} for both $\mathbf{G}_{c}[A]$ and $\mathbf{L}[A]$, representations given as functional integrations over $A_{\mu}^{a}(x)$-dependence that is not more complicated than Gaussian.

Now, due to the non-Abelian nature of the theory, the gauge field dependence of the causal Green's function in these Fradkin representations is contained within an ordered exponential (OE), as described in the Appendix,
\begin{eqnarray}
\left( e^{ -ig \int_{0}^{s}{ds' \, u'_{\mu}(s') \, A_{\mu}^{a}(y-u(s')) \, \tau^{a}} + g \int_{0}^{s}{ds' \sigma_{\mu \nu} \, \mathbf{F}_{\mu \nu}^{a}(y-u(s')) \, \tau^{a}}} \right)_{+},
\end{eqnarray}

\noindent where $u_{\mu}(s')$ is the relevant Fradkin variable~\cite{HMF1990}, and where the $( \cdots)_+$-prescription is to mean ordering in $s'$. Note that for the sake of simplicity, the spin-related term was absent in our previous analysis~\cite{Fried2009}.  To extract the $A_{\mu}^{a}(x)$-dependence out of the ordered exponential, one can introduce two functional integrals for delta-functionals relevant to the first and second term in the ordered exponential; and the two delta-functionals can themselves be further represented by two sets of functional integrals. Proceeding this way, the quark Green's function can be written as
\begin{eqnarray}\label{Eq:GreenFunctionFullExpansion01}
& & \mathbf{G}_{c}(x, y | A) \\ \nonumber &=&  i \mathcal{N}_{\Omega} \, \mathcal{N}_{\Phi} \, \int_{0}^{\infty}{ds \, \int{d[u]} \, \int{d[\alpha] \, \int{d[\mathbf{\Xi}] \, \int{d[\Omega] \, \int{d[\mathbf{\Phi}] \, e^{-is m^{2}}} \, e^{- \frac{1}{2} \Tr{\ln{\left( 2h \right)}} } }}}} \\ \nonumber && \quad \times \delta^{(4)}(x - y + u(s)) \, e^{ \frac{i}{4} \int_{0}^{s}{ds' \, [u'(s')]^{2} } } \\ \nonumber & & \quad \times  e^{-i \int_{0}^{s}{ds' \, \Omega^{a}(s') \, \alpha^{a}(s')} - i \int_{0}^{s}{ds' \, \mathbf{\Phi}^{a}_{\mu \nu}(s') \,  \mathbf{\Xi}^{a}_{\mu \nu}(s') } } \, \left( e^{ i \int_{0}^{s}{ds' \, \left[ \alpha^{a}(s') - i \sigma_{\mu \nu} \, \mathbf{\Xi}_{\mu \nu}^{a}(s') \right] \, \tau^{a}}} \right)_{+}  \\ \nonumber & & \quad \times {\left[ m - \gamma_{\mu} \frac{\delta}{\delta u'_{\mu}(s)} \right]} \\ \nonumber & & \quad \times e^{- i g \int{d^{4}z \, \left[ 2 \left(\partial_{\nu} \mathbf{\Phi}^{a}_{\nu \mu}(z) \right) + \int_{0}^{s}{ds' \, \delta^{(4)}(z-y+u(s')) \, u'_{\mu}(s') \, \Omega^{a}(s') } \right] \, A^{a}_{\mu}(z) }}  \\ \nonumber & & \quad \times e^{ + i g^{2} \int{d^{4}z \, f^{abc} \, \mathbf{\Phi}^{a}_{\mu \nu}(z) \, A^{b}_{\mu}(z) \, A^{c}_{\nu}(z) }},
\end{eqnarray}

\noindent where the detailed derivation is given in the Appendix. One sees that the A-dependence in both G[A] and L[A] is no worse than Gaussian, and hence the linkage operation can be carried through exactly.

\par
What this means is that the sum of all relevant Feynman graphs of arbitrary complication can be obtained exactly, and the result expressed in terms of Fradkin's representations. And since Fradkin's representations are Potential Theory constructs, they are not difficult to approximate in almost any physical situation, and especially at high energies, as in the eikonal representation used in Ref.~\cite{Fried2009}.

\section{\label{Sec:4}Effective Locality}

To display the {{Effective Locality}} property of all such QCD correlation functions, combine the Gaussian $A$-dependence of every $\mathbf{G}_{c}[A]$ entering the process into the quantity
\begin{equation}
\exp{\left[ \frac{i}{2} \int{\mathrm{d}^{4}z  \, A^{a}_{\mu}(z) \, \mathcal{K}^{ab}_{\mu \nu}(z) \, A^{b}_{\nu}(z) } + i \int{\mathrm{d}^{4}z \, \mathcal{Q}^{a}_{\mu}(z) A^{a}_{\mu}(z)} \right]},
\end{equation}

\noindent where $\mathcal{K}$ and $\mathcal{Q}$ are local functions of the Fradkin variables, collectively denoted by $u_{\mu}(s_{i})$, the $\Omega^{a}(s_{i})$ and the $\Phi^{a}_{\mu \nu}(s_{i})$ needed to extract the $A^{a}_{\mu}(y - u(s'))$ from an ordered exponential.  Note that $\mathcal{K}$ and $\mathcal{Q}$ will also represent the sum of similar contributions from each of the $\mathbf{G}_{c}(x, y |A)$ which collectively generate the amplitude under consideration. For example, in the case of  the 4-point function, one will get
\begin{eqnarray}\label{Eq:K04}
\mathcal{K}_{\mu \nu}^{a b}(z) = &+& 2 g^{2} \int_{0}^{s}{ds' \, \delta^{(4)}(z-y_{\mathrm{I}}+u(s')) \, f^{abc} \mathbf{\Phi}^{c}_{\mathrm{I},\mu \nu}(s')}  \\ \nonumber &+& 2 g^{2} \int_{0}^{\bar{s}}{d\bar{s}' \, \delta^{(4)}(z-y_{\mathrm{I\!I}}+\bar{u}(\bar{s}')) \, f^{abc} \mathbf{\Phi}^{c}_{\mathrm{I\!I},\mu \nu}(\bar{s}')}
\end{eqnarray}

\noindent and
\begin{eqnarray}\label{Eq:Q04}
\mathcal{Q}_{\mu}^{a}(z) = &-& 2g \, \partial_{\nu} \mathbf{\Phi}^{a}_{\mathrm{I},\nu \mu}(z) - g \int_{0}^{s}{ds' \, \delta^{(4)}(z-y_{\mathrm{I}}+u(s')) \, u'_{\mu}(s') \Omega^{a}_{\mathrm{I}}(s') }  \\ \nonumber &-& 2 g \, \partial_{\nu} \mathbf{\Phi}^{a}_{\mathrm{I\!I},\nu \mu}(z) - g \int_{0}^{\bar{s}}{d\bar{s}' \, \delta^{(4)}(z-y_{\mathrm{I\!I}}+\bar{u}(\bar{s}')) \, \bar{u}'_{\mu}(\bar{s}') \, \Omega^{a}_{\mathrm{I\!I}}(\bar{s}') },
\end{eqnarray}

\noindent where the subscripts $1,2$ and $\mathrm{I},\mathrm{I\!I}$ are used (interchangeably) to denote particles $\mathrm{I}$ and $\mathrm{I\!I}$; and, for the purpose of tracking, the barred variables are used to denote the particle $\mathrm{I\!I}$.  Similarly, the notation
\begin{equation}
\mathbf{\Phi}^{a}_{\mu \nu}(z) \equiv \int_{0}^{s}{ds' \, \delta^{(4)}(z - y + u(s')) \, \mathbf{\Phi}^{a}_{\mu \nu}(s')}
\end{equation}

\noindent has been introduced in ${\cal{Q}}$ for ease of presentation. For higher quark n-point functions, there will be additional terms contributing to ${\cal{K}}$ and to ${\cal{Q}}$, but their forms will be the same.

Combining the quadratic and linear A-dependence with $\mathcal{K}$ and $\mathcal{Q}$ above, and that explicitly written in Eq.~(\ref{Eq:1}), the operation needed becomes
\begin{equation}\label{Eq:2}
e^{-\frac{i}{2} \int{\frac{\delta}{\delta A} \cdot \mathbf{D}_{\mathrm{c}}^{(\zeta)} \cdot  \frac{\delta}{\delta A}} } \cdot \left[ e^{+ \frac{i}{2} \int{A \cdot \bar{\mathcal{K}}  \cdot A} + i \int{\bar{\mathcal{Q}} \cdot A}} \cdot e^{\mathbf{L}[A]} \right],
\end{equation}

\noindent where
\begin{equation}\label{Eq:K05}
\bar{\mathcal{K}}_{\mu \nu}^{a b} = \mathcal{K}_{\mu \nu}^{a b}(z) + g f^{abc} \chi_{\mu \nu}^{c}(z) + \left({\mathbf{D}_{\mathrm{c}}^{(\zeta)}}^{-1}\right)_{\mu \nu}^{a b}
\end{equation}

\noindent and
\begin{equation}\label{Eq:Q05}
\bar{\mathcal{Q}}_{\mu}^{a}(z) = \mathcal{Q}_{\mu}^{a}(z) + \partial_{\nu} \chi_{\mu \nu}^{a}(z).
\end{equation}

\noindent In $\bar{\mathcal{K}}$, all terms but the inverse of the gluon propagator are local. Eq.~(\ref{Eq:2}) requires the linkage operator to act upon the product of two functionals of $A$, which we write in the easily-derived form
\begin{equation}\label{Eq:3}
\left. e^{\mathfrak{D}_{A}} \cdot \left( \mathcal{F}_{\mathrm{I}}[A] \, \mathcal{F}_{\mathrm{I\!I}}[A]\right) = \left(e^{\mathfrak{D}_{A}} \cdot \mathcal{F}_{\mathrm{I}}[A] \right) \cdot e^{\overleftrightarrow{\mathfrak{D}}} \cdot  \left(e^{\mathfrak{D}_{A'}} \cdot \mathcal{F}_{\mathrm{I\!I}}[A'] \right) \right|_{A'=A},
\end{equation}

\noindent where, using an obvious notation, the "cross-linkage" operator $e^{\overleftrightarrow{\mathfrak{D}}}$ is defined by
\begin{equation}
\overleftrightarrow{\mathfrak{D}} = -i \int{\overleftarrow{\frac{\delta}{\delta A}} \cdot \mathbf{D}_{\mathrm{c}}^{(\zeta)} \cdot  \overrightarrow{\frac{\delta}{\delta A'}}}.
\end{equation}

\noindent With the identifications,
\begin{equation}
\mathcal{F}_{\mathrm{I}}[A] = \exp{\left[\frac{i}{2} \int{A \cdot \bar{\mathcal{K}}  \cdot A} + i \int{\bar{\mathcal{Q}} \cdot A}\right]}\ ,\ \ \ \mathcal{F}_{\mathrm{I\!I}}[A] = \exp{\left(\mathbf{L}[A] \right)},
\end{equation}
the evaluation of $\left( e^{\mathfrak{D}_{A}} \cdot \mathcal{F}_{\mathrm{I}}[A] \right)$ is given by a standard functional identity~\cite{HMF1972,HMF1990}
\begin{eqnarray}\label{Eq:4}
&& \exp{\left[ \frac{i}{2} \int{\bar{\mathcal{Q}} \cdot \mathbf{D}_{c}^{(\zeta)} \cdot \left(1 - \bar{\mathcal{K}} \cdot \mathbf{D}_{c}^{(\zeta)} \right)^{-1} \cdot \bar{\mathcal{Q}} } + \frac{1}{2} \Tr{\ln{\left(1 -  \bar{\mathcal{K}} \cdot \mathbf{D}_{c}^{(\zeta)} \right) }} \right]} \\ \nonumber & & \cdot \exp{\left[ \frac{i}{2} \int{A \cdot \bar{\mathcal{K}} \cdot \left( 1 - \mathbf{D}_{c}^{(\zeta)} \cdot \bar{\mathcal{K}} \right)^{-1} \cdot A } + \int{\bar{\mathcal{Q}} \cdot \left( 1 - \mathbf{D}_{c}^{(\zeta)} \cdot \bar{\mathcal{K}} \right)^{-1} \cdot A} \right]},
\end{eqnarray}

\noindent where the quantity $\mathbf{D}_{c}^{(\zeta)}\left(1 -  \bar{\mathcal{K}} \cdot \mathbf{D}_{c}^{(\zeta)}\right)^{-1}$ reduces to
\begin{equation}
\mathbf{D}_{c}^{(\zeta)}\left(1 -  \bar{\mathcal{K}} \cdot \mathbf{D}_{c}^{(\zeta)}\right)^{-1}=\mathbf{D}_{c}^{(\zeta)}\left(1 - \hat{\mathcal{K}} \cdot \mathbf{D}_{c}^{(\zeta)} - {\mathbf{D}_{c}^{(\zeta)}}^{-1} \cdot \mathbf{D}_{c}^{(\zeta)}\right)^{-1} = (- \hat{\mathcal{K}})^{-1},
\end{equation}

\noindent where, now, $\hat{\mathcal{K}} = \mathcal{K} + g (f \cdot \chi)$. In the limit $A \rightarrow 0$, Eq.~(\ref{Eq:3}) may then be replaced by
\begin{eqnarray}\label{Eq:5}
&& \exp{\left[ -\frac{i}{2} \int{\bar{\mathcal{Q}} \cdot \hat{\mathcal{K}}^{-1} \cdot \bar{\mathcal{Q}} } + \frac{1}{2} \Tr{\ln{\left( \hat{\mathcal{K}}\right) }} + \frac{1}{2} \Tr{\ln{\left( -\mathbf{D}_{c}^{(\zeta)}\right) }} \right]} \\ \nonumber & & \cdot \exp{\left[ \frac{i}{2} \int{\frac{\delta}{\delta A'} \cdot \mathbf{D}_{c}^{(\zeta)} \cdot \frac{\delta}{\delta A'}}\right]} \cdot \exp{\left[ \frac{i}{2} \int{ \frac{\delta}{\delta A'} \cdot \hat{\mathcal{K}}^{-1} \cdot \frac{\delta}{\delta A'}} - \int{\hat{\mathcal{Q}} \cdot \bar{\mathcal{K}}^{-1} \cdot \frac{\delta}{\delta A'}} \right]},
\end{eqnarray}

\noindent which quantity is now to operate upon $\left(e^{\mathfrak{D}_{A'}} \cdot \mathcal{F}_{\mathrm{I\!I}}[A'] \right)$.

Now observe that the first term on the second line of Eq.~(\ref{Eq:5}) is exactly $\exp{\left[-\mathfrak{D}_{A'} \right]}$, and serves to remove the $\exp{\left[\mathfrak{D}_{A'} \right]}$ of the operation $e^{\mathfrak{D}_{A'}} \cdot \mathcal{F}_{\mathrm{I\!I}}[A']$.  With the exception of an irrelevant $\exp{\left[ \frac{1}{2} \Tr{\ln{\left( -\mathbf{D}_{c}^{(\zeta)}\right)}}\right]}$ factor, to be absorbed into an overall normalization, what remains, to all orders of coupling and for every such process, is the quantity
\begin{equation}\label{Eq:6}
\exp{\left[ -\frac{i}{2} \int{\bar{\mathcal{Q}} \cdot \hat{\mathcal{K}}^{-1} \cdot \bar{\mathcal{Q}} } + \frac{1}{2} \Tr{\ln{\left( \hat{\mathcal{K}}\right) }}\right]}
\end{equation}

\noindent multiplying the operation
\begin{equation}\label{Eq:7}
\exp{\left[\frac{i}{2} \int{\frac{\delta}{\delta A} \cdot \hat{\mathcal{K}}^{-1} \cdot \frac{\delta}{\delta A}}\right]} \cdot \exp{\left[ -\int{\bar{\mathcal{Q}} \cdot \hat{\mathcal{K}}^{-1}  \cdot \frac{\delta}{\delta A}}\right]} \cdot \exp{\mathbf{L}[A]},
\end{equation}

\noindent where the now-useless prime of $A'$ has been suppressed.
\par
It is clear that nothing in Eqs.~(\ref{Eq:6}) and (\ref{Eq:7}) ever refers to $\mathbf{D}_{c}^{(\zeta)}$, which means that gauge-invariance is here rigorously achieved as a matter of gauge-independence.  The extraordinary feature of this result is that, because $\hat{\mathcal{K}} = \mathcal{K} + g (f \cdot \chi)$ and the $\mathcal{K}$ and $\mathcal{Q}$ coming from $\mathbf{L}[A]$ are all local functions, with non-zero matrix elements $\langle x | \hat{\mathcal{K}} | y \rangle =  \hat{\mathcal{K}}(x) \, \delta^{(4)}(x-y)$, the contributions of Eqs.~(\ref{Eq:6}) and (\ref{Eq:7}) will depend only on the Fradkin variables $u(s')$ and the space-time coordinates $y_{i}$ in a specific but local way.
If one expands the exponential of the closed-loop functional in Eq. (\ref{Eq:2}) as
\begin{equation}\label{Eq:Lexpansion01}
e^{\mathbf{L}[A]} = 1 + \mathbf{L}[A] + \frac{1}{2!} \, \mathbf{L}[A] \,\mathbf{L}[A] + \cdots,
\end{equation}
\noindent the exponent in each expansion term is at most quadratic in $A^{a}_{\mu}$, as seen from the $\mathbf{L}[A]$ of the Appendix.  One then collects the linear and quadratic $A^{a}_{\mu}$-dependence of $\exp{\mathbf{L}[A]}$, and to that, adds the Gaussian A-dependence coming from any combination of the $G_c[A]$.  The linkage operation can be performed exactly, and taking the limit $A \rightarrow 0$ yields
\begin{eqnarray}\label{Eq:4b}
&& \exp{\left[ \frac{i}{2} \int{\bar{\mathcal{Q}} \cdot \mathbf{D}_{c}^{(\zeta)} \cdot \left(1 - \bar{\mathcal{K}} \cdot \mathbf{D}_{c}^{(\zeta)} \right)^{-1} \cdot \bar{\mathcal{Q}} } + \frac{1}{2} \Tr{\ln{\left(1 -  \bar{\mathcal{K}} \cdot \mathbf{D}_{c}^{(\zeta)} \right) }} \right]} \\ \nonumber &=& \exp{\left[ -\frac{i}{2} \int{\bar{\mathcal{Q}} \cdot \hat{\mathcal{K}}^{-1} \cdot \bar{\mathcal{Q}} } + \frac{1}{2} \Tr{\ln{\left( \hat{\mathcal{K}}\right) }} + \frac{1}{2} \Tr{\ln{\left( -\mathbf{D}_{c}^{(\zeta)}\right) }} \right]},
\end{eqnarray}

\noindent where $\hat{\mathcal{K}}$ and $\bar{\mathcal{Q}}$ here include terms from the expansion of $\exp{\mathbf{L}[A]}$.  Again, the gauge invariance will be manifest, except for the unimportant factor $\exp{\left[ \frac{1}{2} \Tr{\ln{\left( -\mathbf{D}_{c}^{(\zeta)}\right)}}\right]}$ absorbed in the normalization.

One should note that quite similar, local forms involving $(g f \cdot \chi)^{-1}$ were previously obtained by Reinhardt \emph{et al.}~\cite{Reinhardt1993a} in an instanton approximation to a functional integral over gluon fluctuations.  The present article shows that, in the physical, Minkowskian spacetime, such local dependence is an integral part of the exact theory.

% The reason why such a property has remained unnoticed for decades (with the exception of Ref.\cite{Reinhardt1993a}, but in a fairly different perspective) is that considerations and calculations similar to (\ref{Eq:3}) have been carried out in most customary form of gauge-invariant functional integrals,
% \begin{equation}\label{Eq:8}
% \mathcal{N} \, \int{ \rm{d}[A] \, \delta{\left[\mathcal{F}(A)\right]} \, \det{\left[ \delta{\mathcal{F}}/{\delta \omega}\right]} \, e^{ - \frac{i}{2} \, \int{ A \cdot D^{-1} \cdot A} } \, \mathcal{F}_{\mathrm{I}}(A) \, \mathcal{F}_{\mathrm{I\!I}}(A)}.
% \end{equation}
%
% \noindent Were the Gribov-copies issue kept under control during the course of its evaluation, then Eq.(\ref{Eq:8}) would be equivalent to (\ref{Eq:3}). Still, the effective locality property displayed by the {\it{linkage formalism}} through Eqs. (\ref{Eq:3}) - (\ref{Eq:7}), would not be discernable from (\ref{Eq:8}).

Again, it is important to stress that this result is peculiar to the cubic and quartic gluon interactions of QCD. The manifestly gauge invariant construction of Ref.~\cite{Fried2009} does not work for QED, whereas in QCD it is made possible by the Halpern functional integral over $\chi_{\mu\nu}$.  Effective locality is the statement seen in the linkage formalism, in which Halpern's functional integral is reduced to a finite set of ordinary integrals, for every connected amplitude.

This property makes for an interesting contrast with the non-manifestly gauge-invariant QED, where the {\it{carriers}} of each interaction are the {\it{action-at-a-distance}} $\mathbf{D}_{c}(w-z)$, where $w$ and $z=y-u(s')$ are relevant space-time and Fradkin variables.  Here, in QCD, each $\mathbf{D}_{c}$ is replaced by the local $\hat{\mathcal{K}}^{-1}(w)$ at a particular value of $w$, say $w_{0}$; and this quantity is multiplied by $\delta^{(4)}(w_{0} - z)$.  An immediate consequence is that all values of the $w$ parameter in the original functional integration over $\int{\mathrm{d}[{\chi}(w)]}$ are, with their normalization factors, removed from the problem, replacing functional integrals by an ordinary integral
\begin{equation}
\int{\mathrm{d}^{n}{\chi}(w_{0})}
\end{equation}

\noindent where $n$ denotes the number of independent ${\chi}^{a}_{\mu \nu}$ fields over which one is to integrate.  More generally, for a correlation function describing the interaction of $N$ quarks and/or anti-quarks, the result will be given as a product of no more than $N$ such ordinary integrals, $\prod_{\ell = 1}^{N}{\int{\mathrm{d}^{n}{\chi}(w_{\ell})}}$.  However, since the difference between these $\omega_{\ell}$ is given by the sums and differences of the quark position variables $y_{\ell}$ and the Fradkin $u(s_{\ell})$, there are situations in which the number of such $\omega_{\ell}$ variables can be dramatically reduced.

\section{\label{Sec:5}Discussion and Summary }

One important proviso to all of the above statements should be made, which is most easily seen in the eikonal model of Ref.~\cite{Fried2009}.  The basic formalism we have used really refers to particles, which may be real or virtual, and not to quarks which are always virtual in the sense that they exist asymptotically only in bound states, and not individually.  At some point a modification must be made to take this difference into account. In Ref.~\cite{Fried2009} that change was made by the observation that while the longitudinal momentum and energy of a pair of quarks in their CM could be (roughly) estimated in terms of the bound state in which each exists, that is not possible for the transverse momentum of either quark.  The transverse momentum of a quark within its bound state is, in principle, not a measurable quantity, and as such cannot be specified asymptotically.  In Ref.~\cite{Fried2009} this problem was treated by replacing a transverse $\delta^{(2)}(\vec{y}_{1\perp} - \vec{y}_{2\perp}) = \delta^{(2)}(\vec{b})$ by a peaked
\begin{equation}
\varphi(\vec{b}^{\, 2}) \sim \int{\frac{\mathrm{d}^{2}k}{(2\pi)^{2}} \, e^{i\vec{k} \cdot \vec{b} - \vec{k}^{2}/M^{2}}},
\end{equation}

\noindent where $M \simeq \mathcal{O}(\mbox{total CM scattering energy})$.  This non-unique but simple form was chosen as a convenient way of expressing the statement that small $b$, or the large frequency components of $\vec{k}_{\perp}$ cannot be specified with precision. A specific method of achieving such necessary transverse imprecision will appear in the third paper of this Series.

%
% new add-on from Herb's email
%
It may be useful to give an example of the meaning and power of Effective Locality in the context of the simplest, non-trivial, eikonal model of Ref. \cite{Fried2009}. There, the argument $w$ of the Halpern variable $\widehat{\chi}(w)$ was shown to be fixed at a specific value $w_{0} = (\vec{y}_{T}, 0_{L}; y_{0})$, where $y$ denotes the CM space-time coordinate of one of the scattering quarks or antiquarks. Because of Effective Locality, this is the only ${\chi}(w)$ which is relevant to the interaction; all of the other ${\chi}(w)$, for $w \neq w_{0}$ surrounded by an infinitesimally small 4-volume of amount $(\delta)^{4}$, are automatically removed with their normalization factors from the problem, leaving a single, normalized integral over $\mathrm{d}^{n}{\chi}(w_{0})$.

But what happens when $y$ changes, and $w_{0}$ changes to another $w$-value, say $w_{1}$? The answer is: absolutely nothing, because all the other $w$-values, now including $w_{0}$, will cancel away with their normalizations, replacing the Halpern functional integral by a normalized ordinary integral over $\mathrm{d}^{n}{\chi}(w_{1})$. The ${\chi}(w_{1})$ dependence of the integrand of this new ${\chi}$ integral is exactly the same as that of the integrand centered about ${\chi}(w_{0})$, and must yield exactly the same value as the previous integral over $\mathrm{d}^{n}{\chi}(w_{0})$. In brief, integration over $\mathrm{d}^{n}{\chi}$ is independent of the value of $w$, and this latter coordinate may be suppressed; the Halpern functional integral for this $Q$/$\bar{Q}$ scattering process always reduces to a single $n$-dimensional integral ($n=8$, for SU(3)), which can be evaluated numerically, or approximated by a relevant physical argument. This is an extraordinary simplification to any fundamental calculation, and it is due to Effective Locality.

%

% In a non-eikonal, unquenched QCD calculation, one will find a similar situation, with a $\delta^{(4)}(u(s_{1}) - u(s_{2}))$ appearing  in the
% (exponentiated) radiative corrections of each $\mathbf{G}_{c}[A]$, and a corresponding $\delta^{(4)}(u(s') - v(t') + x' - y)$ in those of $\mathbf{L}[A]$,
% where $x'$ refers to the spatial coordinates of its closed quark loop, and $v(t')$ to its Fradkin variable.
% How to adapt these integrand $\delta$-functions to non-eikonal scattering and production processes is left for another publication.

%

In Summary, the general structure of this "ghost formalism" of non-perturbative QCD should be clear.  It converts the original Halpern functional integral into an "effectively local" theory, where the only remaining  functional integrals are those of the Fradkin representations and integrations over color-changing parameters, both of which categories are amenable to a variety of reasonable approximations, especially at high energies.

% Put \label in argument of \section for cross-referencing
%\section{\label{}}
%\subsection{}
%\subsubsection{}

% If in two-column mode, this environment will change to single-column
% format so that long equations can be displayed. Use
% sparingly.
%\begin{widetext}
% put long equation here
%\end{widetext}

% Specify following sections are appendices. Use \appendix* if there
% only one appendix.
\appendix

\section{\label{Appendix} Modified Fradkin's Represenation of Green's function and Closed-Fermion-Loop Functional}

The causal Green's function in Eq.~(\ref{Eq:GreensFunction01}) can be written as~\cite{HMF1990}
\begin{equation}\label{Eq:GreensFunction03}
\mathbf{G}_{c}[A] = [ m  + i \gamma \cdot \Pi][m  + (\gamma \cdot \Pi)^{2}]^{-1} = [ m  + i \gamma \cdot \Pi] \cdot i \int_{0}^{\infty}{ds \, e^{-ism^{2}} \, e^{is(\gamma \cdot \Pi)^{2}}  },
\end{equation}

\noindent where $\Pi = i [\partial_{\mu} - i g A_{\mu}^{a} \tau^{a}]$ and $(\gamma \cdot \Pi)^{2} = \Pi^{2} + i g \sigma_{\mu \nu} \, \mathbf{F}_{\mu \nu}^{a} \tau^{a}$ with $\sigma_{\mu \nu} = \frac{1}{4} [\gamma_{\mu}, \gamma_{\nu}]$.  Following the Fradkin's method and replacing $\Pi_{\mu}$ with $i \frac{\delta}{\delta v_{\mu}}$, one obtains
\begin{eqnarray}
& & \mathbf{G}_{c}(x,y|A) \\ \nonumber &=& i \int_{0}^{\infty}{ds \ e^{-ism^{2}} \cdot e^{i \int_{0}^{s}{ds'
\frac{\delta^{2}}{\delta v_{\mu}^{2}(s')}}} \cdot \left[ m - \gamma_{\mu} \, \frac{\delta}{\delta v_{\mu}(s)} \right] } \, \delta( x -y + \int_{0}^{s}{ds' \ v(s')}) \\ \nonumber & & \times \left.  \left(\exp{\left\{ -ig \int_{0}^{s}{ds' \left[v_{\mu}(s') \, A_{\mu}^{a}(y-\int_{0}^{s'}{v}) \tau^{a} + i \sigma_{\mu \nu} \, \mathbf{F}_{\mu \nu}^{a}(y-\int_{0}^{s'}{v}) \tau^{a} \right] }\right\}} \right)_{+}  \right|_{v_{\mu} \rightarrow 0}.
\end{eqnarray}

\noindent Then, one can insert~\cite{Fried2002a,HMF2002}
\begin{equation}
1 = \int{\mathrm{d}[u] \, \delta(u(s') - \int_{0}^{s'}{ds'' \ v(s'')})}
\end{equation}

\noindent and replace the delta-functional $\delta(u(s') - \int_{0}^{s'}{ds'' \ v(s'')})$ with a functional integral over $\Omega$, then the Green's function becomes~\cite{YMS2008}
\begin{eqnarray}
& & \mathbf{G}_{c}(x,y|A) \\ \nonumber &=&  i \int_{0}^{\infty}{ds \ e^{-is m^{2}}} \, e^{- \frac{1}{2} \Tr{\ln{\left( 2h \right)}} } \, \int{d[u]} \, e^{ \frac{i}{4} \int_{0}^{s}{ds' \, [u'(s')]^{2} } } \, \delta^{(4)}(x - y + u(s)) \\ \nonumber & & \quad \times {\left[ m + i g \gamma_{\mu} A_{\mu}^{a}(y-u(s)) \tau^{a} \right]} \, \left( e^{ -ig \int_{0}^{s}{ds' \, u'_{\mu}(s') \, A_{\mu}^{a}(y-u(s')) \, \tau^{a}} + g \int_{0}^{s}{ds' \sigma_{\mu \nu} \, \mathbf{F}_{\mu \nu}^{a}(y-u(s')) \, \tau^{a}}} \right)_{+},
\end{eqnarray}

\noindent where $h(s_{1},s_{2})=\int_{0}^{s}{ds' \, \theta(s_{1} - s') \theta(s_{2} - s')}$.  To remove the $A_{\mu}^{a}$-dependence out of the linear (mass) term, one can replace $i g A_{\mu}^{a}(y-u(s)) \tau^{a}$ with $- \frac{\delta}{\delta u'_{\mu}(s)}$ operating on the ordered exponential so that
\begin{eqnarray}
\mathbf{G}_{c}(x,y|A) &=&  i \int_{0}^{\infty}{ds \ e^{-is m^{2}}} \, e^{- \frac{1}{2} \Tr{\ln{\left( 2h \right)}} } \, \int{d[u]} \, e^{ \frac{i}{4} \int_{0}^{s}{ds' \, [u'(s')]^{2} } } \, \delta^{(4)}(x - y + u(s)) \\ \nonumber & & \quad \times {\left[ m - \gamma_{\mu} \frac{\delta}{\delta u'_{\mu}(s)} \right]} \, \left( e^{ -ig \int_{0}^{s}{ds' \, u'_{\mu}(s') \, A_{\mu}^{a}(y-u(s')) \, \tau^{a}} + g \int_{0}^{s}{ds' \sigma_{\mu \nu} \, \mathbf{F}_{\mu \nu}^{a}(y-u(s')) \, \tau^{a}}} \right)_{+}.
\end{eqnarray}

\noindent To extract the $A$-dependence out of the ordered exponential, one may use the identities
\begin{eqnarray}
1 &=& \int{d[\alpha] \, \delta{\left[ \alpha^{a}(s') + g u'_{\mu}(s') \, A^{a}_{\mu}(y-u(s'))\right]}}  \\ \nonumber 1 &=& \int{d[\mathbf{\Xi}] \, \delta{\left[ \mathbf{\Xi}^{a}_{\mu \nu}(s') - g \mathbf{F}_{\mu \nu}^{a}(y-u(s'))\right]} }
\end{eqnarray}

\noindent so that
\begin{eqnarray}
& & \left( e^{ -ig \int_{0}^{s}{ds' \, u'_{\mu}(s') \, A_{\mu}^{a}(y-u(s')) \, \tau^{a}} + g \int_{0}^{s}{ds' \sigma_{\mu \nu} \, \mathbf{F}_{\mu \nu}^{a}(y-u(s')) \, \tau^{a}}} \right)_{+} \\ \nonumber &=& \mathcal{N}_{\Omega} \, \mathcal{N}_{\Phi} \, \int{d[\alpha] \, \int{d[\mathbf{\Xi}] \, \int{d[\Omega] \, \int{d[\mathbf{\Phi}] \, \left( e^{ i \int_{0}^{s}{ds' \, \left[ \alpha^{a}(s') - i \sigma_{\mu \nu} \, \mathbf{\Xi}_{\mu \nu}^{a}(s') \right] \, \tau^{a}}} \right)_{+} }}}} \\ \nonumber & & \quad \times e^{-i \int{ds' \, \Omega^{a}(s') \, \alpha^{a}(s')}  - i \int{ds' \, \mathbf{\Phi}^{a}_{\mu \nu}(s') \,  \mathbf{\Xi}^{a}_{\mu \nu}(s') } } \\ \nonumber & & \quad \times e^{- i g \int{ds' \, u'_{\mu}(s') \, \Omega^{a}(s') \, A^{a}_{\mu}(y-u(s')) } + i g \int{ds' \, \mathbf{\Phi}^{a}_{\mu \nu}(s') \, \mathbf{F}_{\mu \nu}^{a}(y-u(s'))}  },
\end{eqnarray}

\noindent where $\mathcal{N}_{\Omega}$ and $\mathcal{N}_{\Phi}$ are constants which normalize the functional representations of the delta-functionals.  All $A$-dependence is removed from the ordered exponentials and the resulting form of the Green's function is exact (it entails no approximation).  Alternatively, extracting the $A$-dependence out of the ordered exponential can also be achieved by using the functional translation operator. One writes
\begin{equation}
\left( e^{ + g \int_{0}^{s}{ds' \, \left[ \sigma_{\mu \nu} \, \mathbf{F}_{\mu \nu}^{a}(y-u(s')) \tau^{a} \right]} }\right)_{+}  = \left. e^{g \int_{0}^{s}{ds' \, \mathbf{F}_{\mu \nu}^{a}(y-u(s')) \, \frac{\delta}{\delta \mathbf{\Xi}_{\mu\nu}^{a}(s')}}} \cdot \left( e^{\int_{0}^{s}{ds' \, \left[ \sigma_{\mu \nu} \, \mathbf{\Xi}_{\mu \nu}^{a}(s') \tau^{a} \right]} }\right)_{+} \right|_{\mathbf{\Xi} \rightarrow 0}.
\end{equation}

For the closed-fermion-loop functional $\mathbf{L}[A]$, one can write~\cite{HMF1990}
\begin{equation}\label{Eq:ClosedFermionLoopFunctional02}
\mathbf{L}[A] = - \frac{1}{2} \, \int_{0}^{\infty}{\frac{ds}{s} \, e^{-ism^{2}} \, \left\{\Tr{\left[ e^{-is(\gamma \cdot \Pi)^{2}} \right]} - \left\{ g=0 \right\} \right\}},
\end{equation}

\noindent where the trace sums over all degrees of freedom. The Fradkin's representation proceeds along the same steps as in the case of $\mathbf{G}_{c}[A] $, and the closed-fermion-loop functional reads
\begin{eqnarray}\label{Eq:LFradkin01}
\mathbf{L}[A] &=&  - \frac{1}{2} \int_{0}^{\infty}{\frac{ds}{s} \, e^{-is m^{2}}} \, e^{- \frac{1}{2} \Tr{\ln{(2h)}} } \\ \nonumber && \quad \times \int{d[v]} \, \delta^{(4)}(v(s)) \, e^{ \frac{i}{4} \int_{0}^{s}{ds' \, [v'(s')]^{2} } } \\ \nonumber & & \quad \times \int{d^{4}x \, \tr{\left( e^{ -ig \int_{0}^{s}{ds' \, v'_{\mu}(s') \, A_{\mu}^{a}(x-v(s')) \, \tau^{a}} + g \int_{0}^{s}{ds' \sigma_{\mu \nu} \, \mathbf{F}_{\mu \nu}^{a}(x-v(s')) \, \tau^{a}}} \right)_{+}} } \\ \nonumber & & - \left\{ g = 0 \right\},
\end{eqnarray}

\noindent where the trace now sums over color and spinor indices. Also, Fradkin's variables have been denoted by $v(s')$, instead of $u(s')$, in order to distinguish them from those appearing in the Green's function $\mathbf{G}_{c}[A] $. One finds
\begin{eqnarray}\label{Eq:LFradkin02}
\mathbf{L}[A] &=&  - \frac{1}{2} \int_{0}^{\infty}{\frac{ds}{s} \, e^{-is m^{2}}} \, e^{- \frac{1}{2} \Tr{\ln{(2h)}} } \\ \nonumber && \quad \times \mathcal{N}_{\Omega} \, \mathcal{N}_{\Phi} \int{d^{4}x \, \int{\mathrm{d}[\alpha] \, \int{\mathrm{d}[\Omega] \, \int{\mathrm{d}[\mathbf{\Xi}] \, \int{\mathrm{d}[\mathbf{\Phi}] \, }}}}}  \\ \nonumber & & \quad \times \int{d[v] \, \delta^{(4)}(v(s)) \, e^{ \frac{i}{4} \int_{0}^{s}{ds' \, [v'(s')]^{2} } }  } \\ \nonumber & & \quad \times \ e^{-i \int{ds' \, \Omega^{a}(s') \, \alpha^{a}(s')} - i \int{ds' \, \mathbf{\Phi}^{a}_{\mu \nu}(s') \,  \mathbf{\Xi}^{a}_{\mu \nu}(s') } }  \cdot  \tr{\left( e^{ i \int_{0}^{s}{ds' \, \left[ \alpha^{a}(s') - i \sigma_{\mu \nu} \, \mathbf{\Xi}_{\mu \nu}^{a}(s') \right] \, \tau^{a}}} \right)_{+}} \\ \nonumber & & \quad \times  e^{- i g \int_{0}^{s}{ds' \, v'_{\mu}(s') \, \Omega^{a}(s') \, A^{a}_{\mu}(x - v(s'))} - 2 i g \int{d^{4}z \, \left(\partial_{\nu}  \mathbf{\Phi}^{a}_{\nu \mu}(z) \right) \, A^{a}_{\mu}(z) }} \\ \nonumber & & \quad \times e^{ + i g^{2} \int{ds' \, f^{abc} \mathbf{\Phi}^{a}_{\mu \nu}(s') \, A^{b}_{\mu}(x- v(s')) \, A^{c}_{\nu}(x- v(s')) }}  \\ \nonumber & & - \left\{ g = 0 \right\},
\end{eqnarray}

\noindent where the same properties as those of $\mathbf{G}_{c}[A] $ can be read off explicitly and the $A$-dependence is at most Gaussian.

%
%
%

% If you have acknowledgments, this puts in the proper section head.
%\begin{acknowledgments}
%HMF is supported in part by a Julian Schwinger Foundation Travel Grant.
%\end{acknowledgments}

% Create the reference section using BibTeX:
\bibliography{QCD_Locality_bib}

\end{document}